\RequirePackage{fix-cm}
\documentclass[smallextended]{svjour3}       
\smartqed  
\usepackage{graphicx}
\usepackage{graphics}
\usepackage{amssymb}
\begin{document}

\title{Singular Schr\"odinger operators and Robin billiards
}
\subtitle{Spectral properties and asymptotic expansions}


\author{Pavel Exner
}


\institute{P. Exner \at
              Department of Theoretical Physics, Nuclear Physics Institute \\
              Czech Academy of Sciences, 25068 \v Re\v z near Prague, Czechia, and \\
              Doppler Institute for Mathematical Physics and Applied Mathematics
              \\ Czech Technical University, B\v rehov\'a 7, 11519 Prague, Czechia \\
              Tel.: +420-776-154-823\\
              \email{exner@ujf.cas.cz}           
}

\date{Received: date / Accepted: date}

\maketitle

\begin{abstract}
This paper summarizes the contents of a plenary talk at the Pan African Congress of Mathematics held in Rabat in July 2017. We provide a survey of recent results on spectral properties of Schr\"odinger operators with singular interactions supported by manifolds of codimension one and of Robin billiards with the focus on the geometrically induced discrete spectrum and its asymptotic expansions in term of the model parameters.
\end{abstract}

\section{Introduction}
\label{intro}

In this paper we will investigate several classes of problems. Most of them are related with singular Schr\"odinger operators that can be formally written as
\begin{equation}\label{leakyH}
H_{\alpha,\Gamma}= -\Delta -\alpha \delta (x-\Gamma )\,,\quad \alpha>0\,,
\end{equation}
in $L^2(\mathbb{R}^d)$, where the support $\Gamma$ is a set of Lebesgue measure zero and some geometric properties, for instance, a curve or a metric graph in $\mathbb{R}^2$, a surface in $\mathbb{R}^3$, etc.

The motivation is twofold. Viewed from the mathematician's ivory tower, they are simply interesting objects in which different areas like operator theory and PDE on the one hand, and geometry and topology of the set $\Gamma$ on the other hand, are related in a nontrivial way. But there is also a practical side of this effort coming from the need to model a large class of nanostructures usually dubbed \emph{quantum graphs} and similar objects. The models commonly used and widely studied \cite{BK13} have one drawback, namely that they neglect quantum tunneling between different part of such a structure. Operators of the type (\ref{leakyH}) represent an alternative for which the term \emph{leaky quantum graphs} is often used -- cf., e.g., \cite{Ex08} or \cite[Chap.~10]{EK15}.

The topic is rather wide and we will adopt some limitation in the present paper. In particular, we will focus on interaction supports of \emph{codimension one} and on singular potentials of the $\delta$ type, higher codimensions and more singular interactions will be mentioned only episodically.

One the other hand, we are also going to discuss another class of problems which might be regarded as `one-sided' analogues of the operator (\ref{leakyH}) describing motion in regions $\subset \mathbb{R}^d$ with Robin (i.e., mixed, or third-type) boundary conditions. We will see that the one-sided character makes these systems in many respects different from the earlier mentioned class.

In all cases our interest concerns primarily the \emph{discrete spectrum} of the operators involved, its existence and \emph{asymptotic expansions} of the eigenvalues in terms the parameters of the model. Let us describe the contents of the paper. In the next section we will set the scene with a proper definition of the operators involved, and we will also collect basic facts about their spectra. Section~\ref{sec:strong} is devoted to asymptotic behavior of the discrete spectrum in the situation when the singular interaction in (\ref{leakyH}) is \emph{strongly attractive}, $\alpha\to\infty$, the analogous asymptotic problem for Robin billiards is discussed in Section~\ref{sec:robin}. In Section~\ref{sec:geometric} we turn to a different sort of expansions, this time connected with weak \emph{geo\-metric perturbations} of the trivial supports $\Gamma$ such as a line in the plane or a plane in $\mathbb{R}^3$. In Section~\ref{sec:magnetic} we return to billiards, this time in the particular shape of a planar wedge, and amend them with a homogeneous \emph{magnetic field}; we will investigate the existence of the discrete spectrum. We conclude the paper with remarks about open problems. Since this is a survey, proofs of our results will be only hinted, however, references will be always given to sources where they can be found in their entirety.

\section{Preliminaries}
\label{sec:prelim}

\subsection{Definition of the operator}
\label{sec:definition}

Let us look first how one can give meaning to the heuristic expression. If $\Gamma$ is a \emph{smooth manifold} with $\mathrm{codim}\,\Gamma=1$ one can define a self-adjoint operator, $H_{\alpha,\Gamma}$, or alternatively\footnote{Both the symbols are used in the literature for the same operator.} $-\Delta_{\delta,\alpha}$, by changing the domain of the Laplacian: we suppose that the operator acts as $-\Delta$ on functions from $H_\mathrm{loc}^2 (\mathbb{R}^d \setminus \Gamma)$, which are continuous at $\Gamma$ and exhibit there a normal-derivative jump,
\begin{equation}\label{bc1}
\left.\frac{\partial \psi }{\partial n}(x)\right|_+
-\left.\frac{\partial \psi }{\partial n }(x)\right|_- =-\alpha\psi (x)\,.
\end{equation}
This is a physicist's notation with the derivative taken in a fixed normal direction at the point $x\in\Gamma$ which corresponds well to the formal expression as describing the attractive  $\delta$-interaction of strength $\alpha$ perpendicular to $\Gamma$ at the point $x\:$ \cite[Chap.~I.3]{AGHH}. A mathematician would prefer to take the \emph{sum} of the derivatives with respect to the outward derivatives and to stress that the objects entering relation (\ref{bc1}) are traces of the functions involved \cite[Rem.~2.9]{BEL14}.

The drawback of such a definition is that we want to have the operator also defined for less regular sets $\Gamma$. It is easy to check that the above introduced operator $H_{\alpha,\Gamma}$ is associated with the quadratic form
\begin{equation}\label{form1}
q_{\delta,\alpha}[\psi] := \| \nabla \psi\|^2_{L^2(\mathbb{R}^d)} -\alpha \|f|_{\Gamma}\|^2_{L^2(\Gamma)}
\end{equation}
with the domain $H^1(\mathbb{R}^d)$. This makes it possible to introduce a substantially wider class of operators. We start from the following definition:

\smallskip

\noindent A finite family of Lipschitz domains $\mathcal{P} = \{\Omega_k\}_{k=1}^n$ is called a \emph{Lipschitz partition} of $\mathbb{R}^d$, $\,d\geq 2$, if
$$
\mathbb{R}^d = \bigcup_{k=1}^n\overline\Omega_k \quad \mathrm{and} \quad \Omega_k\cap\Omega_l = \emptyset\,, \quad k,l=1,2,\dots,n,\;\;k\ne l\,.
$$
The union $\cup_{k=1}^n\partial\Omega_k =: \Gamma$ is the \emph{boundary} of  $\mathcal{P}$. For $k\not= l$ we set $\Gamma_{kl}:= \partial\Omega_k\cap\partial\Omega_l$ and we say that $\Omega_k$ and $\Omega_l$,  $k\ne l$, are neighboring domains if $\sigma_{k}(\Gamma_{kl})>0$, where $\sigma_k$ is the Lebesgue measure on $\partial\Omega_k$.

\smallskip

Using these notions we can state the following result \cite{BEL14}:
\begin{proposition}
Let $\mathcal{P} = \{\Omega_k\}_{k=1}^n$ be a Lipschitz partition of $\mathbb{R}^d$ with the boundary $\Gamma$, and let $\alpha:\Gamma\rightarrow\mathbb{R}$ belong to $\ L^\infty(\Gamma)$. Then the quadratic form $q_{\delta,\alpha}$ defined above is closed and semibounded from below.
\end{proposition}

Consequently, there is unique self-adjoint operator $H_{\alpha,\Gamma}$, or $-\Delta_{\delta,\alpha}$, associated with this quadratic form. Note that this definition is not only more general in term of the support regularity but also allows for the coupling strength varying along $\Gamma$. In particular, the interaction support may be in fact a \emph{proper subset} of $\Gamma$, since $\alpha$ may vanish on a part of this set. We will use this fact to define $H_{\alpha,\Gamma}$ on sets $\Gamma$ with a boundary, otherwise we will always assume in the following that $\alpha$ \emph{is constant on its `true' support}.

\begin{remark} \label{delta'}
Similarly one can define more singular interactions on sets of codimension one. Prominent among them is the $\delta'$ interaction which on a smooth $\Gamma$ is characterized by a modification of the boundary condition (\ref{bc1}), namely the continuity of the normal derivative and a jump of the function value, $\psi|_+-\psi|_-=\beta\partial_n\psi$; for less regular supports one can again employ quadratic forms \cite{BEL14}. In a similar way one can also define the general four-parameter family of singular interactions supported by $\Gamma\,$ \cite{ER16}.
\end{remark}

\begin{remark} \label{codim2}
A more singular problem is also represented by supports of higher codimensions. It follows from general principles that they can be constructed provided $\mathrm{codim\,}\Gamma\le 3$. Most attention has been paid in the literature to the case $\mathrm{codim\,}\Gamma=2$ where the operator can be defined via boundary conditions matching generalized boundary values \cite{EKo02}.
\end{remark}

\subsection{Spectrum of $-\Delta_{\delta,\alpha}$}
\label{sec:spectrum}

In general, the spectrum is determined both by the \emph{geometry of $\Gamma$} and the coupling $\alpha$, in particular, by its \emph{sign}. If $\Gamma$ is \emph{compact}, it is easy to see that $\sigma_\mathrm{ess}(-\Delta_{\delta,\alpha}) = \mathbb{R}_+$. On the other hand, the essential spectrum may change if the support $\Gamma$ is non-compact. As an example, take a line in the plane and suppose that $\alpha$ is \emph{constant and positive}; by separation of variables one finds easily that in this case $\sigma_\mathrm{ess}(-\Delta_{\delta,\alpha}) = [-\frac14 \alpha^2, \infty)$.

The question about the \emph{discrete spectrum} is more involved. Suppose first that the interaction support is  \emph{finite}, $|\Gamma|<\infty$. It is clear from the above claim about the essential spectrum and (\ref{form1}) that $\sigma_\mathrm{disc}(-\Delta_{\delta,\alpha}) =\emptyset$ holds if the interaction is \emph{repulsive}, $\alpha\le 0$. For an attractive coupling, on the other hand, the negative discrete spectrum may be non-empty, but whether it is the case is determined by the dimension. Specifically, for $d=2$ bound states exist whenever $|\Gamma|>0$, in particular, we have a weak-coupling expansion \cite{KL14}
 $$
\lambda(\alpha) = \big(C_\Gamma+o(1)\big)\, \exp\left(- \frac{4\pi}{\alpha|\Gamma|} \right) \quad \mathrm{as}\quad \alpha|\Gamma| \to 0+
 $$
This is not the case for $d=3$ where the singular coupling must exceed a critical value to bind. As an example, let $\Gamma$ be a sphere of radius $R>0$ in $\mathbb{R}^3$, then by \cite{AGS87} we have
 $$
 \sigma_\mathrm{disc}(H_{\alpha,\Gamma}) \ne \emptyset \quad \mathit{iff}\quad \alpha R>1\,,
 $$
and in the same way, weakly bound states do not exist in dimensions $d>3$.

\begin{remark} \label{otherweak}
For the more singular interactions mentioned in the above remarks the weak coupling behavior may look differently. A curve in $\mathbb{R}^3$ may have no bound states if it too short or the interaction is too weak \cite{EKo08}. The $\delta'$ interaction supported by a planar loop is more interesting as the result depends here on the \emph{topology} of the support. If $\Gamma$ is a loop a discrete spectrum is nonempty as a simple variational argument shows, on the other hand, a \emph{non-closed} curve has no bound states if the coupling is weak enough \cite{JL16}.
\end{remark}

\subsection{Geometrically induced bound states}
\label{sec:induced}

The above results may seem predictable because $H_{\alpha,\Gamma}$ is after all nothing but a Schr\"odinger operator. What could be more surprising is that the geometry itself may induce a discrete spectrum. As an example, consider the case $d=2$ and suppose that $\Gamma$ is an infinite curve. We have mentioned above that if the latter is a line, the spectrum is purely essential, $\sigma(-\Delta_{\delta,\alpha}) = [-\frac14 \alpha^2, \infty)$. Let us now look what a geometric perturbation could do. To be specific, we consider a \emph{non-straight, piecewise $C^1$-smooth curve} $\Gamma:\: \mathbb{R} \to \mathbb{R}^2$ parameterized by its arc length\footnote{With an abuse of notation, we employ here and in the following the same symbol for the interaction support as a set as well as for the function that parametrizes it.}, $|\Gamma(s) - \Gamma(s')| \le |s-s'|$, assuming in addition that
\begin{itemize}
\setlength{\itemsep}{2pt}
\item $|\Gamma(s) - \Gamma(s')| \ge c|s-s'|$ holds for some $c\in(0,1)$
\item $\Gamma$ is \emph{asymptotically straight}: there are $d>0$, $\:\mu> \frac{1}{2}$ and $\omega\in(0,1)$ such that
$$
1-\, {|\Gamma(s)-\Gamma(s')|\over|s-s'|} \le d \left\lbrack
1+|s+s'|^{2\mu} \right\rbrack^{-1/2}
$$
holds in the sector $S_\omega:= \left\{ (s,s'):\: \omega < {s\over s'} < \omega^{-1}\, \right\}$.
\end{itemize}
Under these assumptions we have the following result \cite{EI01}:
 \begin{theorem} \label{thm:geom}
$\sigma_\mathrm{ess}(-\Delta_{\delta,\alpha}) = [-\frac{1}{4}\alpha^2,\infty)$ and, in addition, $-\Delta_{\delta,\alpha}$ has at least one eigenvalue in $(-\infty,-\frac{1}{4}\alpha^2]$.
 \end{theorem}
\emph{Sketch of the proof:} The argument employs the (generalized) Birman-Schwinger principle \cite[Thm.~6.7]{EK15} by which an eigenvalue $-\zeta^2$ with $\zeta>0$ of $-\Delta_{\delta,\alpha}$ exists \emph{iff} $\mathcal{R}^\zeta_{\alpha,\Gamma}\psi = \psi$ where $\mathcal{R}^\zeta_{\alpha,\Gamma}$ is integral operator on $L^2(\mathbb{R})$ with kernel
$$
\mathcal{R}^\zeta_{\alpha,\Gamma}(s,s') := \frac{\alpha}{2\pi} K_0(\zeta|\Gamma(s)-\Gamma(s')|)\,.
$$
The bending is regarded as a \emph{perturbation of the straight line} for which the equation $\mathcal{R}^\zeta_{\alpha,\Gamma}\psi = \psi$ is of a convolution type and the spectrum of $\mathcal{R}^\zeta_{\alpha,\Gamma}$ is easily found to be $\big[0,\frac{\alpha}{2\zeta}\big]$. The crucial observation is that, in view of the free resolvent kernel properties, this perturbation is \emph{sign definite}, and furthermore in view of our asymptotic straightness assumption it is \emph{compact}. Thus the spectrum of $\mathcal{R}^\zeta_{\alpha,\Gamma}$ for the non-straight $\Gamma$ extends above $\frac{\alpha}{2\zeta}$ but the added part may consist of isolated eigenvalues only. It is easy to check that those depend continuously on $\zeta$ and tend to zero as $\zeta\to\infty$, hence there is a value $\zeta>\frac{\alpha}{2\pi}$ at which such a curve crosses the value one. \hfill $\Box$

\medskip

Geometrically induced bound states exist in other situations too, let us briefly recall some presently known results:
\begin{itemize}
\setlength{\itemsep}{2pt}
\item the above result and its extensions mentioned below have implications for \emph{more complicated Lipschitz partitions}: if $\tilde\Gamma \supset \Gamma$ holds in the set sense, then we have $H_{\alpha, \tilde\Gamma} \le H_{\alpha, \Gamma}$. If the essential spectrum thresholds are the same -- which is often easy to establish -- then $\sigma_\mathrm{disc}(H_{\alpha,\tilde \Gamma})\ne\emptyset$ holds whenever the same is true for $\sigma_\mathrm{disc}(H_{\alpha,\Gamma})$
\item in \emph{higher dimensions} the situation is more complicated. For \emph{smooth curved surfaces},  $\Gamma\subset\mathbb{R}^3$, an analogous existence result is proved in the strong coupling asymptotic regime, $\alpha\to\infty$, only \cite{EKo03}
\item on the other hand, we can mention the example of a \emph{conical surface} of an opening angle $\theta\in(0,\frac12\pi)$ in $\mathbb{R}^3$, where for any constant $\alpha>0$ we have $\sigma_\mathrm{ess}(-\Delta_{\delta,\alpha}) = \mathbb{R}_+$ and an \emph{infinite numbers of eigenvalues below $-\frac14 \alpha^2$} accumulating at the threshold \cite{BEL14}
\item moreover, the above result remains valid for any \emph{local deformation} of the conical surface. We also know the accumulation rate for conical layers:  by \cite{LO16} it is
$$
\mathcal{N}_{-\textstyle{\frac14}\alpha^2-E}(-\Delta_{\delta,\alpha}) \sim \frac{\cot\theta}{4\pi}\, |\ln E|\,,\quad E\to 0+\,,
$$
and a similar result also holds for \emph{non-cylindrical} cones \cite{OP18}
\item on the other hand, the result is again dimension-dependent: for a conical surface in $\mathbb{R}^d,\: d>3$, we have $\sigma_\mathrm{disc}(-\Delta_{\delta,\alpha})=\emptyset$, cf. \cite{LO16}
\item for Schr\"odinger operators with $\delta'$ interaction mentioned in Remark~\ref{delta'} above one can often establish the existence of geometrically induced bound states comparing them to $-\Delta_{\delta,\alpha}$ with the same $\Gamma$ and $\alpha:=\frac4\beta$, cf.~\cite{BEL14}
\end{itemize}

\section{Strong coupling asymptotics}
\label{sec:strong}

Our next topic is the \emph{strong-coupling behavior} of the discrete spectrum. The main idea is that for large values of $\alpha$ the eigenfunctions are localized transversally being concentrated in the vicinity of $\Gamma$ and the problem becomes effectively $(d-1)$-dimensional; the question is how are the geometric properties of the support manifested. We have to adopt stronger regularity assumption.  Consider first a $C^4$ smooth \emph{curve in $\mathbb{R}^2$ without ends and self-intersections}, either infinite or a closed loop. In the limit $\alpha\to\infty$ the $j$-th eigenvalue of $H_{\alpha,\Gamma}$ behaves as
\begin{equation}\label{asym2D}
\lambda_j(\alpha) = -\frac14\,\alpha^2 + \mu_j +
\mathcal{O}(\alpha^{-1} \ln\alpha)\,,
\end{equation}
where $\mu_j$ is the $j$-th eigenvalue of the operator
\begin{equation}\label{compar2D}
S_\Gamma = -{\mathrm{d}\over \mathrm{d}s^2} - {1\over 4}\,\kappa(s)^2
\end{equation}
on $L^2(I)$ where $I$ refers to the arc-length parameter being either $\mathbb{R}$ or a finite interval with periodic boundary conditions and $\kappa(s)$ is the signed curvature of $\Gamma$ at the point $s$, cf.~\cite{EY02} or \cite[Thm.~4.1]{Ex08}.

Under similar hypotheses on \emph{smoothness} and \emph{absence of boundaries} imposed on a \emph{surface in $\mathbb{R}^3$}, we have according to \cite{EKo03} the same asymptotic expansion, (\ref{asym2D}), however, $\mu_j$ is now the $j$-th eigenvalue of
\begin{equation}\label{compar3D}
S_\Gamma = -\Delta_\Gamma + K - M^2 \,,
\end{equation}
where $-\Delta_\Gamma$ is the Laplace-Beltrami operator on $\Gamma$ and $K, M$, respectively, are the corresponding \emph{Gauss} and \emph{mean} curvatures of the surface.

Let us recall the technique which allows one to derive these results. It has three essential ingredients. The first is \emph{Dirichlet-Neumann bracketing} \cite[Sec.~XIII.15]{RS}: we impose additional boundary conditions at the boundary $\Sigma_a$ of the tubular neighborhood of $\Gamma$ of the halfwidth $a$. This yields a two-sided bound on $-\Delta_{\delta,\alpha}$ and we have only to care about the neighborhood part because we are interested in the negative spectrum and the Dirichlet/Neumann Laplacian in the remaining part of $\mathbb{R}^d$ is positive.

In the second step one uses inside the tubular neighborhood the natural curvilinear coordinates, sometimes named after Fermi, and estimates the coefficients to squeeze the operator between those with \emph{separated variables}. For a curve in $\mathbb{R}^2$, e.g., they are
$$
\tilde{H}^{\pm}_{a,\alpha} =U^{\pm}_{a}\otimes 1+1\otimes T^{\pm}_{a,\alpha}\,,
$$
where
$$
U^{\pm}_{a}=-(1\mp a\|\kappa\|_\infty)^{-2} \frac{\mathrm{d}^{2}}
{\mathrm{d}s^2}+V_{\pm}(s)
$$
with periodic b.c. in the case of a loop, where $V_-(s)\le -\frac14 \kappa^2(s) \le V_+(s)$ with an $\mathcal{O}(a)$ error. In other words, the $U^{\pm}_{a}$ are \emph{$\mathcal{O}(a)$ close to $S_\Gamma$}. The transverse operators, on the other hand, are associated with the forms
$$
t^{+}_{a,\alpha}[f]=
\int^{a}_{-a}|f^{\prime}(u)|^{2}\,\mathrm{d}u-\alpha |f(0)|^{2}
$$
and $t^{-}_{a,\alpha}[f]= t^{+}_{a,\alpha}[f] - \|k\|_\infty(|f(a)|^2 +|f(-a)|^2)$ defined on $H^1_0(-a,a)$ and $H^1(-a,a)$, respectively. We observe that for large values of $\alpha$ the presence of the boundaries causes just an exponentially small error:
 \begin{lemma}
There is a positive $c_N$ such that $T_{\alpha,a}^{\pm}$ has for $\alpha$ large enough a single negative eigenvalue, being a negative square of $\zeta_{\alpha,a}^{\pm}$ satisfying
$$
-\frac{\alpha ^{2}}{4} \left(1+c_N
\,\mathrm{e}^{-\alpha a/2} \right) <\zeta_{\alpha,a}^{-}
< -\frac{\alpha ^{2}}{4}< \zeta_{\alpha,a}^{+} < -\frac{\alpha
^{2}}{4} \left(1 -8 \,\mathrm{e}^{-\alpha a/2} \right)
$$
 \end{lemma}

\smallskip

\noindent In the final step we relate the neighborhood halfwidth with the coupling constant choosing $a= 6\alpha^{-1} \ln\alpha$ which yields the result. In the dimension three the argument proceeds in the similar way, the only difference is that we cannot `straighten' the layer neighborhood fully, the geometry of surface $\Gamma$ remains to be present in the Laplace-Beltrami operator \cite{EKo03}.

\smallskip

\noindent The technique sketched above works in a number of other situations:
\begin{itemize}
\setlength{\itemsep}{2pt}
\item instead of a loop or an asymptotically straight curve one can consider \emph{periodic manifolds}, connected or disconnected; the asymptotic expansion analogous to (\ref{asym2D}) can be then derived for the fibre (Bloch) eigenvalues, cf.~\cite[Secs.~4.3 and 4.4]{Ex08} for details
\item in a similar way one can treat two-dimensional \emph{loops in a magnetic field} perpendicular to the plane, where the role of the quasimomentum is played by the magnetic flux through the loop, cf.~\cite[Sec.~4.5]{Ex08}
 \item one can also derive asymptotic expansions for interactions supported by curves in $\mathbb{R}^3$ mentioned in Remark~\ref{codim2}. In that case the divergent first term in (\ref{asym2D}) is replaced by $4\,\mathrm{e}^{2(-2\pi\alpha+\psi(1))}$ where $-\psi(1)\approx 0.577$ is Euler-Mascheroni constant, i.e. the eigenvalue of a two-dimensional point interaction, cf.~\cite[Chap.~I.5]{AGHH} and \cite[Thm.~4.2]{Ex08}; recall that the strong coupling regime means in this case the limit $\alpha\to-\infty$
 \item with a small modification one can also derive in this way asymptotic expansions for the $\delta'$ interactions mentioned in Remark~\ref{delta'}, cf.~\cite{EJ13,EJ14}. The divergent term in this case replaced by $-\frac{4}{\beta^2}$ and the error term is $\mathcal{O}(\beta|\ln\beta|)$, recall that weak coupling here means the limit $\beta\to 0$
\end{itemize}

On the other hand, the described technique is of limited use in the situation when $\Gamma$ has a boundary as, for instance, a \emph{finite} curve in the plane. The reason comes from the Dirichlet and Neumann conditions which now have to be imposed also on the `lids' of the tubular neighborhood $\Gamma_a$. As a result we get an estimate in which the comparison operator (\ref{compar2D}) appears with different boundary conditions which is sufficient to estimate asymptotically the \emph{number of eigenvalues} but it is too rough to pinpoint each of them separately. A natural conjecture is that the `correct' boundary conditions for (\ref{compar2D}) are in this respect \emph{Dirichlet}. It appears that it indeed the case \cite{EP14}:

\begin{theorem}
Suppose that $\Gamma$ is a $C^4$ smooth open arc in $\mathbb{R}^2$ of length $L$ with \emph{regular ends}, i.e. there is neighborhood in which the curve can smoothly extended. Then the strong-coupling expansion of the $j$-th negative eigenvalue of $H_{\alpha, \Gamma}$ is
\begin{equation}\label{arc}
\lambda_j(\alpha)=-\frac14\, \alpha^2 +\mu_j +\mathcal{O}\Big(\frac{\ln\alpha}{\alpha}\Big) \quad \mathrm{as} \quad \alpha\to \infty\,,
\end{equation}
where $\mu_j$ is the $j$-th eigenvalue of operator (\ref{compar2D}) on $L^2(0,L)$ with Dirichlet b.c.
\end{theorem}
\emph{Sketch of the proof:} We employ again a bracketing. The upper (Dirichlet) bound works as before, while for the lower (Neumann) we use the fact that $\Gamma$ has by assumption regular ends. This allows us to take an \emph{`extended' tubular neighbourhood}, at each endpoint longer by $a:=\frac6\alpha \ln\alpha$. The trouble is that now we loose the advantage of variable separation and the task is to show that the Neumann condition imposed at this distance from the curve endpoints will have an effect which can be included into the error term.

The way to find such an estimate presented in \cite{EP14} is based on the (generalized) Birman-Schwinger principle mentioned in the proof of Theorem~\ref{thm:geom} above. It says, in particular, that the eigenfunction of $H_{\alpha,\Gamma}$ corresponding to an eigenvalue $\lambda_j= -\zeta_j^2$ can be written as
 $$
\psi_j(x) = \frac{1}{2\pi} \int_\Gamma K_0(\zeta_j|x-\Gamma(s)|)\, f_j(s)\, \mathrm{d}s\,,
 $$
where $f_j$ is the corresponding eigenfunction of the Birman-Schwinger operator acting on $L^2(\Gamma,\mathrm{d}s) \sim L^2(0,L)$; the claim of the theorem then follows from simple geometric estimates combined with the exponential decay of the Macdonald function $K_0$ at large distances. \hfill $\Box$

In a similar vein one can treat \emph{surfaces with a boundary}. Let $\Gamma\subset\mathbb{R}^3$ be a $C^4$-smooth relatively compact orientable surface with a compact Lipschitz boundary $\partial\Gamma$. In addition, we suppose that $\Gamma$ can be extended through the boundary, i.e. that there exists a larger $C^4$-smooth surface $\Gamma_2$ such that $\overline{\Gamma}\subset \Gamma_2$. As in the case when the boundary is absent we consider the operator $S_\Gamma = -\Delta^D_\Gamma + K - M^2$, where $-\Delta^D_\Gamma$ is Laplace-Beltrami operator on $\Gamma$, now with \emph{Dirichlet condition} at $\partial\Gamma$, and $K, M$, respectively, are again the corresponding \emph{Gauss} and \emph{mean} curvatures.

We denote eigenvalues of this operator as $\mu^D_j,\, j\in\mathbb{N}$, then we have \cite{DEKP16}:

\begin{theorem}
Let $\Gamma$ be as above, then for a fixed $j\in\mathbb{N}$ the asymptotic expansion
$$
\lambda_j (H_{\alpha,\Gamma})= -\frac14\,\alpha^2 + \mu^D_j + o(1) \quad \mathrm{as} \quad \alpha\to\infty
$$
is valid. If, in addition, $\Gamma$ has a $C^2$ boundary, then the remainder estimate can be replaced by $\mathcal{O}\big(\alpha^{-1} \ln\alpha\big)$.
\end{theorem}
\emph{Sketch of the proof:} As in the previous case, the upper bound is easy because one can take a layer neighborhood of the surface $\Gamma$ itself and impose the `correct', that is, Dirichlet conditions at its boundary. Using then an estimate with separated variables, we get the result. The lower bound can be done in two different ways. One is to construct an explicit family of operators, cf.~\cite{DEKP16} for details, using the projection to the lowest transverse mode and its orthogonal complement, and to employ its monotonicity to prove the convergence. This gives the result but without an explicit error term; the advantage is that it requires the Lipshitz property for $\partial\Gamma$ only. An alternative is to use the same idea as for the curves with ends based on Birman-Schwinger principle. This yields an error term, but since the boundary is a more complicated object now, we have to require a $C^2$ smoothness in order to be able to perform the needed geometric estimates. \hfill $\Box$

\section{Robin billiards}
\label{sec:robin}

Let us now pass to our second topic, the motion in a finite region $\subset\mathbb{R}^d$ with a mixed boundary conditions imposed on the boundary, for the sake of brevity we speak of `billiards' with Robin boundary. We start with the two-dimensional situation. Let thus $\Omega$ be an open, simply connected set in $\mathbb{R}^2$ with a closed $C^4$ Jordan boundary $\partial \Omega=\Gamma:[0,L]\ni s\mapsto (\Gamma_1,\Gamma_2)\in\mathbb{R}^2$, with $\kappa: [0,L]\rightarrow\mathbb{R}$ being the signed curvature of $\Gamma$. We consider the boundary-value problem
\begin{equation}\label{Robin}
-\Delta f=\lambda f \;\; \mathrm {in}\;\Omega\,, \quad
\frac{\partial f}{\partial n}=\beta f\;\; \mathrm{on}\;\Gamma\,,
\end{equation}
with $\beta>0,$ where $\frac{\partial}{\partial n}$ is the outward normal derivative. The corresponding self-adjoint operator $H_{\beta}$ is associated with the quadratic form
\begin{equation}\label{Robform}
q_{\beta}[f] =\|\nabla f\|^2_{L^2(\Omega)}-\beta\int\limits_{\Gamma}|f(x)|^2\mathrm{d}s
\end{equation}
defined on $\mathrm{Dom}(q_{\beta}) = H^{1}(\Omega)$.

The spectrum of $H_{\beta}$ is purely discrete, as before we are interested in the behavior of the eigenvalues as $\beta\to\infty$. We consider again the operator $S:= -\frac{\mathrm{d}^2}{\mathrm{d}s^2} -\frac{1}{4}\kappa^2(s)$ on $L^2(0,L)$ with periodic b.c., and furthermore, we introduce the symbols $\kappa^*=\max\limits_{[0,L]}\kappa(s)$ and $\kappa_*=\min\limits_{[0,L]}\kappa(s)$. It may seem that the quadratic forms (\ref{form1}) and (\ref{Robform}) are closely similar to each other and that the previous results might be easily translated to the previous situation. However, more caution is needed. A naive use of the technique that led to the expansion (\ref{asym2D}) yields only a much weaker result,
$$ 
-\Big(\beta+\frac{\kappa^*}{2}\Big)^2+\mu_n+ \mathcal{O}({\small \frac{\log\beta}{\beta}}) \le \lambda_n(\beta)
\le -\Big(\beta+\frac{\kappa_*}{2}\Big)^2+\mu_n+ \mathcal{O}({\small \frac{\log\beta}{\beta}})\,.
$$ 
The reason is that passing to the curvilinear coordinates in the vicinity of the boundary we get in the one-sided case a boundary term containing $\kappa$. If we want estimates with separated variables we have to employ rough bounds with $\kappa^*$ and $\kappa_*$. However, the lower bound can be improved by a variational technique \cite{EMP14}; this yields at least the first two terms in the expansion:

\begin{theorem}
In the asymptotic regime $\beta\rightarrow \infty$ the $j$-th eigenvalue behaves as
$$
\lambda_j(\beta) = -\beta^2-\kappa^*\beta+\mathcal{O}\big(\beta^{2/3}\big)\,.
$$
\end{theorem}
This result can be further improved in several directions, in particular, one can extend it to higher dimensions \cite{PP16}:

\begin{theorem} \label{thm:HK}
Let $H_\beta$ be the Robin Laplacian in an open, connected domain $\Omega \subset\mathbb{R}^d$, $\,d\ge 2$. Its $j$-th eigenvalue behaves in the limit $\beta \to\infty$ as
$$
\lambda_j(\beta) = -\beta^2+ E_j(-\Delta_S -\beta(d-1)H) +\mathcal{O}\big(\log\beta\big)\,,
$$
where the second term is the $j$-th eigenvalue of the indicated operator; here $-\Delta_S$ is the Laplace-Beltrami operator on $S:= \partial\Omega$ and $H$ is the mean curvature of the boundary, $(d-1)H = \kappa_1+\cdots+\kappa_{d-1}$. In particular, if $\Omega$ has a compact $C^2$ boundary, then the second term is given by the maximum of $H$,
$$
\lambda_j(\beta) = -\beta^2 -\beta(d-1)H_\mathrm{max} +o(\beta)\,.
$$
\end{theorem}

\noindent The error term can be further improved if $\partial\Omega$ is more regular \cite{PP16} but it still does not distinguish between individual eigenvalues. This result also illustrates well the \emph{difference between the one- and two-sided situation}. Take $d=3$, then in the Robin case the next-to-leading-order term is linear in the coupling parameter and  the `effective potential' is given by the \emph{mean curvature} only, while for Schr\"odinger operators considered in the previous sections the next-to-leading term is independent of the coupling and the potential is a combination of \emph{Gauss} and \emph{squared mean} curvatures, $K-M^2$.

In the two-dimensional situation, the asymptotic expansion can be under stronger assumptions improved to pinpoint individual eigenvalues \cite{HK17}:
\begin{theorem}
Consider $\Omega\subset\mathbb{R}^2$ with a $C^\infty$ smooth boundary, possibly infinite. Suppose that the curvature $\kappa$ attains its maximum $\kappa_\mathrm{max}$ at a unique point, and the maximum is non-degenerate, i.e. $k_2:= -\kappa''(0)>0$. Then for any positive $j$ there exists a sequence $\{\zeta_{i,j}\}$ such that, for any positive $M$, the $j$-th eigenvalue has for $\beta\to\infty$ the following asymptotic expansion
$$
\lambda_j(\beta) = -\beta^2-\kappa_\mathrm{max}\beta+ (2j-1) \sqrt{\frac{k_2}{2}} |\beta|^{1/2} + \sum_{i=0}^M \zeta_{i,j} |\beta|^{\frac{1-i}{4}} + |\beta|^{\frac{1-M}{4}} o(1)\,.
$$
\end{theorem}

\smallskip

Let us now make a short detour and show that a part of these results can be extended to \emph{nonlinear} eigenvalue problems, specifically to the question about the spectral bottom of the \emph{$p$-Laplacian with Robin boundary conditions}, that is
$$
-\Delta_p u= \Lambda |u|^{p-2} u \quad \mathrm{in}\;\,\Omega\,, \quad |\nabla u|^{p-2} \frac{\partial u}{\partial n} = \beta |u|^{p-2}u \quad \mathrm{at }\;\,\partial\Omega\,,
$$
where $\Delta_p$ is the $p$-Laplacian, $\Delta_p u= \nabla\cdot(|\nabla u|^{p-2} \nabla u)$, and $n$ is the outer unit normal. We ask about the smallest $\Lambda$ satisfying the above equation, i.e.
$$
\Lambda(\Omega,p,\beta) := \inf_{0\ne u\in W^{1,p}(\Omega)} \frac{\int_{\Omega} |\nabla u|^p\,\mathrm{d} x - \beta \int_{\partial\Omega} |u|^p\,\mathrm{d}\sigma}{\int_{\Omega} |u|^p\,\mathrm{d} x}\,.
$$
We call a domain $\Omega\subset\mathbb{R}^d\,,\; d\ge 2$, admissible if
\begin{itemize}
\setlength{\itemsep}{1pt}
\item the boundary $\partial\Omega$ is $C^{1,1}$, i.e. is locally the graph of a function with a Lipschitz gradient
\item the principal curvatures of $\partial\Omega$ are essentially bounded
\item the map $\partial\Omega \times (0,\delta) \ni (s,t)\mapsto s-tn(s) \in \{x\in\Omega:\: \mathrm{dist}(x,\partial\Omega) < \delta\}$ is bijective for some $\delta>0$
\end{itemize}
The mean curvature $H$ of $\partial\Omega$ is, as above, the arithmetic mean of the principal curvatures, and we set $H_\mathrm{max} \equiv H_\mathrm{max}(\Omega) := \sup\mathrm{ess}\, H$. Then we have the following result \cite{KP17}:

\begin{theorem}
For any admissible domain $\Omega\subset\mathbb{R}^d\,,\; d\ge 2$ and any $p\in(1,\infty)$ we have
$$
\Lambda(\Omega,p,\beta) = -(p-1)\beta^{p/(p-1)} -\beta(\nu-1)H_\mathrm{max}(\Omega) +o(\beta)
$$
as $\beta\to\infty$.
\end{theorem}

\smallskip

\begin{remark} \label{positive}
Before proceeding further, let us stress that these asymptotic results hold only for the `attractive' boundary condition, i.e. $\beta>0$ in (\ref{Robform}). In the `repulsive' case, $\beta\to -\infty$, the behavior is completely different \cite{Fi17}: the eigenvalues approach those of the Dirichlet problem in $\Omega$ and the second term in the expansion reflects the behavior of the corresponding Dirichlet eigenfunctions on the entire boundary.
\end{remark}

\section{Asymptotic expansions for geometric perturbations}
\label{sec:geometric}

Returning now to singular Schr\"odinger operators, we note that the strong coupling regime is not the only asymptotic problem the leaky structure can offer. Let us turn to \emph{geometric perturbations}. The simplest example is a \emph{broken line} $\Gamma=\Gamma_\beta$, that is, two halflines meeting at the angle $\pi-\beta$ with a small $\beta>0$. Since $\alpha>0$ is fixed now, we drop it from the notation of $H_{\alpha,\Gamma_\beta} \equiv -\Delta_{\alpha,\Gamma_\beta}$ writing it simply as $H_{\Gamma_\beta}$, etc. By Theorem~\ref{thm:geom} this operator has eigenvalues, in fact a single one for small enough $\beta$, and by a simple scaling argument together with an analogy with bent Dirichlet tubes \cite[Chap.~6]{EK15} lead us to conjecture that
$$
\lambda (H_{\Gamma_\beta}) =  -\frac14{\alpha^2} +a\beta^4 + o(\beta^4)
$$
with some $a<0$ as $\beta\to 0+$. The question now is (a) what is the coefficient $a$, and (b) \emph{whether a similar formula holds for more general slightly bent curves}.

Let us first specify the class of curves we shall consider: $\Gamma$ will be a continuous and piecewise $C^2$ infinite planar curve without self-intersections parametrized by its arc length, i.e. the graph of a piecewise $C^2$-smooth  function $\Gamma:\, \mathbb{R} \to \mathbb{R}^2 $ such that $|\dot\Gamma(s)|=1$. Moreover, we suppose that
\begin{itemize}
\setlength{\itemsep}{1pt}
\item there exists a number $c\in (0,1)$ such that $|\Gamma (s)-\Gamma (s')|\geq c|s-s'|$ holds for $s,s' \in \mathbb{R}$,
\item there are real numbers $s_1 >s_2 $ and straight lines $\Sigma_i,\: i=1,2,$ such that $\Gamma$ coincides with $\Sigma_1$ for  $s\geq s_1$ and with $\Sigma_2$ for $s\leq s_2$,
\item one-sided limits of $\dot\Gamma$ exists at the points where the function $\ddot\gamma$ is discontinuous.
\end{itemize}
In particular, the signed curvature $\kappa(s)=\dot\Gamma_2(s)\ddot\Gamma_1(s) - \dot\Gamma_1(s)\ddot\Gamma_2(s)$ is piecewise continuous and the one-sided limits of $\dot\Gamma$, i.e. \emph{tangent vectors} to the curve at the points of discontinuity exist. We denote them as $\Pi = \{p_i \}_{i=1}^{\sharp \Pi}$ and shall speak of them as of vertices. Consequently, $\Gamma $ consists of  $\sharp \Pi +1$ simple arcs or \emph{edges}, each having as its endpoints one or two of the vertices. The curvature integral describes \emph{bending} of the curve. Specifically, the angle between the tangents at the points $\Gamma(s)$ and $\Gamma (s')$ equals
 $$ 
  \phi (s,s') = \sum _{p_i\in(s,s')} c(p_i)
  +\int_{(s,s')\setminus\Pi} \kappa(\eta)\,\mathrm{d}\eta\,,
 $$ 
where $c(p_i)\in (0,\pi)$ is the exterior angle of the two adjacent edges of $\Gamma $ meeting at the vertex $p_i$. Alternatively, we can understand $\phi (s,s')$ as the integral over the interval $(s,s')$ of $\tilde{\kappa}:\: \tilde{\kappa}(s) = \kappa(s) + \sum_{p\in\Pi} c(p)\,\delta(s-p)$. By assumption $\kappa,\,\tilde{\kappa}$ are compactly supported, thus $\phi(s,s')$ has the same value for all $s<s_2$ and $s_1<s'$ which we shall call the \emph{total bending}. What is important, one can reconstruct $\Gamma$ from $\tilde{k}$, uniquely up to Euclidean transformations, by
 $$ 
  \Gamma (s) = \left( \int_{0}^s \cos \phi (u, 0 )\,\mathrm{d}u
  \,,\int_{0}^s \sin \phi (u, 0 )\,\mathrm{d}u \right)\,.
 $$ 

Now we introduce the one-parameter family of `scaled' curves $\Gamma_\beta$,
 $$ 
  \Gamma _{\beta} (s) = \left(\int_{0}^s \cos \beta \phi (u, 0)\,\mathrm{d}u
  \,,\int_{0}^s \sin \beta \phi (u, 0))\,\mathrm{d}u \right)\,, \quad |\beta|\in(0,1]\,;
 $$ 
note that depending on (non)vanishing of the total bending of $\Gamma$ the limit $\beta\to 0+$ may have a different meaning, which one might characterize as `straightening' or `flattening', respectively. To state the main result, we define next an integral operator $A:\, L^2 (\mathbb{R})\to L^2 (\mathbb{R})$ through its kernel,
 $$ 
 \mathcal{A}(s,s'):= \frac{\alpha^4}{32\pi} K_0'\left( \frac{\alpha }{2}|s-s'|
 \right)\left( |s-s'|^{-1} \left( \int_{s'}^s  \phi\,\mathrm{d}\eta \right)^2 -
 \int_{s'}^s \phi^2\,\mathrm{d}\eta \right)\,,
 $$ 
which has the following property \cite{EKo15}:
\begin{lemma} 
Under the stated assumptions, we have $\int_{\mathbb{R}\times\mathbb{R}}\mathcal{A}(s,s')\,\mathrm{d}s \,\mathrm{d}s '< \infty$.
\end{lemma}

Now we are in position to state the weak-perturbation result \cite{EKo15}:

\begin{theorem}
There is a $\beta_0 >0$ such that for any $\beta \in (-\beta_0,0) \cup (0,\beta_0 )$ the operator $H_{\Gamma_\beta }$ has a unique eigenvalue
$\lambda (H_{\Gamma_\beta })$ which admits the asymptotic expansion
 \begin{equation}\label{eq-asymbeta}
\lambda (H_{\Gamma_\beta })= -\frac14\,\alpha^2 -\left(\int_{\mathbb{R}\times \mathbb{R}}\mathcal{A}(s,s')\,\mathrm{d}s\,\mathrm{d}s' \right)^2 \beta^4 +o(\beta^4)\,.
 \end{equation}
\end{theorem}
\emph{Sketch of the proof:} We refine the technique used in the proof of Theorem~\ref{thm:geom} and employ again the generalized Birman-Schwinger principle noting that not only $-\zeta^2$ with $\zeta>\frac12\alpha$ belongs to $\sigma_\mathrm{disc}(H_{\Gamma_\beta})$ \emph{iff} $\mathrm{ker}(I-\mathcal{R}^\zeta_{ \Gamma_\beta})\ne\emptyset$ but also the dimensions of the latter and of $\mathrm{ker}(H_{\Gamma_\beta}+\zeta^2)$ are the same. The formula (\ref{eq-asymbeta}) is then obtained by analyzing tge spectral behavior of $\mathcal{R}^\zeta_{\Gamma_\beta}$ under the perturbation; the above lemma ensures that the second term of the expansion makes sense. \hfill $\Box$

\begin{example} \label{broken}
Let us return to the \emph{broken-line example} mentioned in the opening to this section. In this case $\mathcal{A}(s,s')$ can be found easily, it vanishes if $s,s'$ have the same sign, being otherwise
$$ 
\mathcal{A}(s,s')= \frac{\alpha^4}{32\pi} K_0'\left( \frac{\alpha }{2}|s-s'|\right)\frac{|ss'|}{|s-s'|}\,\chi_{\Omega}(s,s')\,,
$$ 
where $\chi_{\Omega}(\cdot ,\cdot)$ is the characteristic function of the set $\Omega$, the union of the second and fourth quadrant. The integral of $\mathcal{A}(s,s')$ over the both variable can be computed explicitly giving
 \begin{equation}\label{brline}
\frac{-\frac14\alpha^2 - \lambda (H_{\Gamma_\beta})}{-\frac14\alpha^2} = - \frac{1}{9\pi^2}\beta^4 + o(\beta^4)\,.
\end{equation}
\end{example}

If we go one dimension higher the problem becomes more subtle because then \emph{global properties} of the interaction support play now role; recall that if $\Gamma$ is a (possibly locally deformed) conical surface in $\mathbb{R}^3$ the discrete spectrum is infinite however `flat' the cone may be \cite{BEL14,OP18}. Let us thus restrict our attention to \emph{locally deformed planes}: consider $\Gamma_\beta = \Gamma_\beta(f)\subset\mathbb{R}^3$ with $\beta>0$ given by
$$
\Gamma_\beta :=
\big\{(x_1,x_2,x_3)\in\mathbb{R}^3\colon x_3 = \beta f (x_1, x_2)\big\}\subset \mathbb{R}^3\,,
$$
where $f\colon \:\mathbb{R}^2\to \mathbb{R}$ is a nonzero $C^2$-smooth, compactly supported function and ask how the spectrum of $H_{\beta}$ behaves in the asymptotic regime, $\beta\to 0+$. The answer, obtained again using Birman-Schwinger analysis \cite{EKL18}, brings to mind the weak-coupling behavior of two-dimensional Schr\"odinger operators.

\begin{theorem}
    Let $\alpha > 0$ be fixed and set
    $$
        \mathcal{D}_{\alpha,f} := \int_{\mathbb{R}^2}
        |p|^2\left (\alpha^2 - \frac{2\alpha^3}{\sqrt{4|p|^2 + \alpha^2} + \alpha}\right )|\hat f(p)|^2 \mathrm{d}p > 0\,,
    $$
    where $\hat f$ is the Fourier transform of $f$.
    Then $\#\sigma_\mathrm{disc}(H_{\beta}) = 1$ holds for all sufficiently small $\beta > 0$,
    and moreover, $\lambda(H_{\Gamma_\beta})$ admits the
    asymptotic expansion
    $$
        \lambda(H_{\Gamma_\beta}) =
        -\frac14\,\alpha^2 -
        \exp\left (-\frac{16\pi}{\mathcal{D}_{\alpha,f}\beta^2}\right ) \big (1+o(1)\big)\,,\quad \beta\to 0+\,.
    $$
\end{theorem}

\section{Magnetic wedges}
\label{sec:magnetic}

This section is devoted to a different problem, the existence of bound states in the presence of a magnetic field with a particular geometry. To be specific, consider a charged particle \emph{constrained to a wedge} $\Omega_\phi$ of a opening angle $\phi\in(0,\pi)$ with Neumann or Robin boundary and subject to a \emph{homogeneous magnetic field} perpendicular to the plane. The \emph{scale invariance} of the wedge allows us to assume without loss of generality that the field intensity $B=1$. This problem is of physical interest appearing as a model in the analysis of the
Ginzburg-Landau equation in the regime of \emph{superconductivity onset} in a surface, occurring when the intensity of an exterior magnetic field
decreases from a large, critical, value, cf.~\cite{Ja01,Po15,Ra17} and references therein.

Let us first introduce the operator of interest. To begin with, we employ the circular gauge choosing the vector potential $A \colon\:
\mathbb{R}^2\to\mathbb{R}^2$ as
\[
    A(x_1,x_2)
    =
    \frac12 (-x_2, x_1)^\top\,,\quad x = (x_1,x_2)\in\mathbb{R}^2\,,
\]
and define the associated magnetic gradient, $\nabla_A := i\nabla + A$, and the magnetic first-order Sobolev space $H^1_A(\Omega)$. On the latter we define the (closed, densely defined, and semi-bounded) quadratic form
\begin{equation}\label{mgform}
    H^1_{A}(\Omega)\ni u\mapsto
    \mathfrak{h}[u]:=
    \left\|\nabla_A u\right\|^2_{L^2(\Omega;\mathbb{C}^2)}
    -
    \beta  \| u|_{\Gamma} \|^2_{L^2(\Gamma)}\,,
\end{equation}
which is a magnetic generalization of (\ref{Robform}). To keep track of the parameters, we denote $\mathfrak{h}$  for $\Omega=\Omega_\phi$ by $\mathfrak{h}_{R,\phi,\beta}$. It is useful to consider at the same time $\Omega = \mathbb{R}^2$ where (\ref{mgform}) is a magnetic generalization of (\ref{form1}) when we will write $\mathfrak{h}_{\delta,\phi,\beta}$; the analogous symbols will employed for the associated self-adjoint operators.
In particular, $H_{N,\phi}:= H_{R,\phi,0}$ is the \emph{magnetic Neumann Laplacian} on the wedge $\Omega_\phi$, and $H_{\delta,\phi,\beta}$ is the
\emph{magnetic Schr\"odinger operator} with a $\delta$-interaction supported on the broken line $\Gamma$ supported on the broken line $\Gamma$ (in the notation of previous section we would write $\Gamma=\Gamma_{\pi-\phi}\,$).

Let us first recall what is known and expected:
\begin{itemize}
 \setlength{\itemsep}{2pt}
 \item the quantities $\Theta_{R,\beta} := \inf \sigma_\mathrm{ess}(H_{R,\phi,\beta})$ and $\Theta_{\delta,\beta} := \inf \sigma_\mathrm{ess} (H_{\delta,\phi,\beta})$ are independent of the aperture $\phi$
 \item in the Neumann case we have $\Theta_0 := \Theta_{N,\beta} \simeq 0.5901$
 \item there is a \emph{conjecture} \cite{Ra17} saying that $\sigma_\mathrm{disc}(H_{N,\phi}) \cap (0,\Theta_0) \ne \emptyset$ holds for any aperture $\phi\in(0,\pi)$
 \item this conjecture is proven for $\phi\le\frac12\pi$, cf.~\cite{Ra17}, the unpublished thesis \cite{Bo03} extends this result to apertures $\phi\lesssim 0.511\pi$
 \item in the \emph{nonmagnetic case} a Neumann wedge obviously has \emph{empty discrete spectrum}, in contrast, \emph{Robin wedges} and \emph{leaky broken lines} have bound states, see \cite{EM14} and Theorem~\ref{thm:geom} above
\end{itemize}

While we are not able to prove the above conjecture, our aim is to enlarge the range of parameters in which the answer to the question is affirmative. The method we employ is variational. First we take a simple trial function
$$
    u(r,\theta) = \mathrm{e}^{-ar^2/2}
    \exp\left (i c r \left[\mathrm{e}^{\theta} - \mathrm{e}^{\phi - \theta}\right ]\right ),
    \quad a,c > 0\,,
$$
expressed in the polar coordinates $(r,\theta)$; in contrast to earlier studies it contains the angular-dependent coefficient in the imaginary exponent. Evaluating the quadratic form for fixed $\phi \in (0,\pi)$ and $\beta \in\mathbb{R}$ we arrive, cf.~\cite{ELP17}, at the fourth-order polynomial $P_{\phi,\beta}(x)$,
\begin{equation}\label{easycond}
P_{\phi,\beta}(x) := x^4\left(2\phi - \pi \tanh\left (\textstyle{\frac12}\phi\right ) \right) - 8\Theta_{R,\beta}\phi x^2 - 16\beta\sqrt{\pi} x + 8\phi\,;
\end{equation}
if $\min_{x \in (0,\infty)} P_{\phi,\beta}(x) < 0$, then $\sigma_\mathrm{disc}(H_{R,\phi, \beta}) \cap (-\infty,\Theta_{R,\beta}) \ne \emptyset$ holds. Plotting the $(\phi,\beta)$-plane we get a graphical solution to the condition shown in Fig.~\ref{fig:robin}. Note that this yields the existence of a bound state below the threshold of the essential spectrum also for the repulsive boundary, $\beta<0$ with small absolute value, which \emph{cannot happen without the magnetic field}. Furthermore, one can make the following conclusion:
\begin{figure}
  \includegraphics[width=0.75\textwidth]{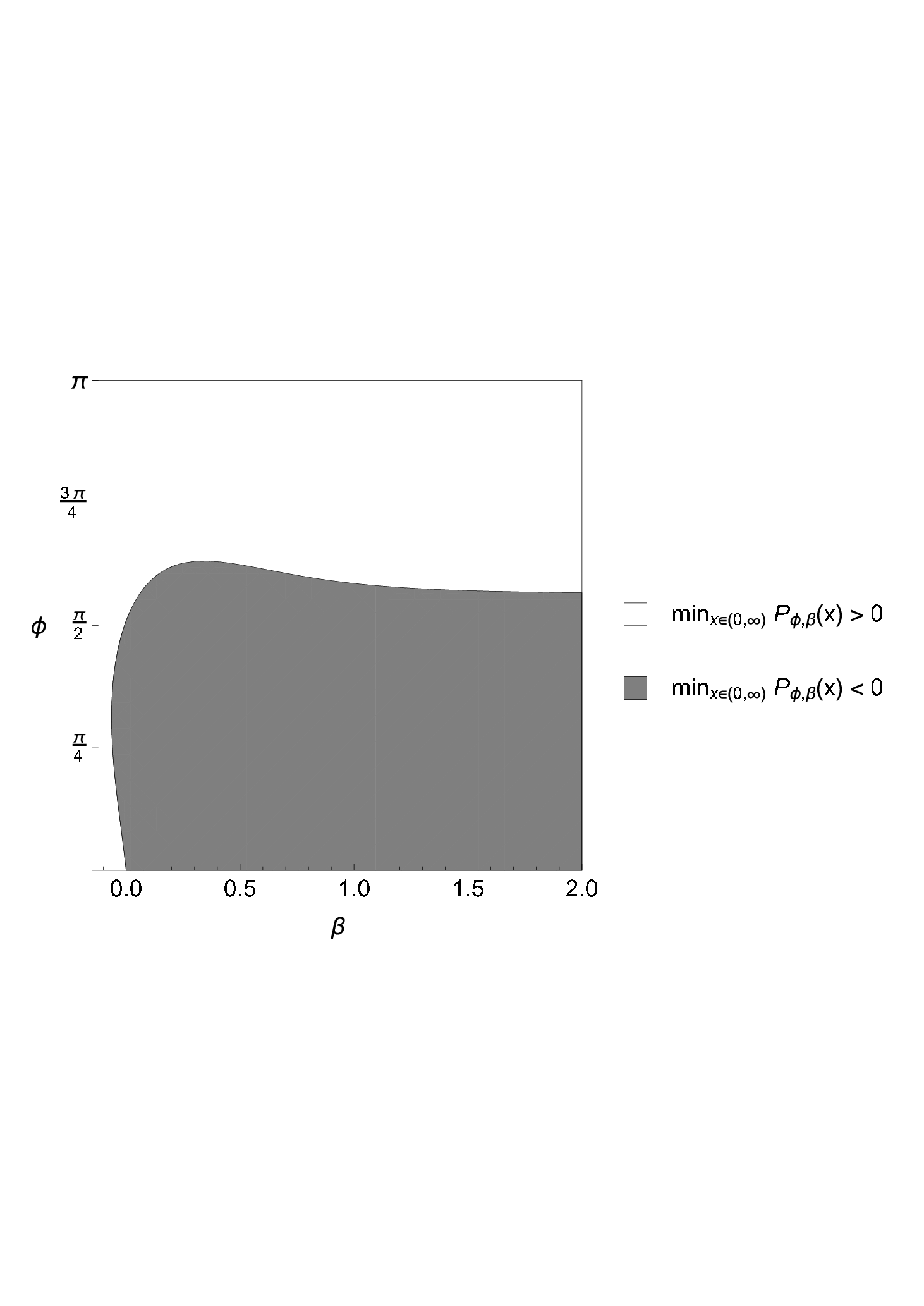}
\caption{The discrete spectrum of the Robin wedge is nonempty in the shaded region}
\label{fig:robin}       
\end{figure}
\begin{proposition}\label{cor2}
    For any wedge aperture $\phi \in (0,\sqrt{\pi})$, that is, $\phi \lesssim
    0.564\pi$,
    $\:\sigma_\mathrm{disc}(H_{R,\phi, \beta}) \cap (-\infty,\Theta_{R,\beta}) \ne
    \emptyset$
    holds  for all $\beta > 0$ large enough.
\end{proposition}

From the physical point of view, however, the Neumann case, $\beta = 0$, is the most interesting, and there the indicated variational argument yields the sufficient condition $\phi\lesssim 0.509\pi$, slightly worse than
$\lesssim 0.511\pi$ obtained in \cite{Bo03}.

To get a better result we use a more sophisticated trial function,
$$
u_\star(r,\theta) = e^{-ar^2/2} \exp\left (i \sum_{k=1}^N r^k b_k(\theta)\right ), \quad N\in\mathbb{N}\,,
$$
with the parameter $a > 0$ and arbitrary real-valued functions $b_k\in C^\infty([0,\phi])$, $\:k=1,2,\dots,N$. Using functional derivative we can get an optimal choice of $\{b_k\}_{k=1}^N$ by solving an appropriate system of linear
second-order ordinary differential equations on the interval $[0,\phi]$ with constant coefficients. In particular, for $N=2$ we arrive at the following conclusion \cite{ELP17}:

\begin{theorem}
Let $s := \sqrt{9-2\pi}$, $\mu_{1,2} := \frac{s \pm 1}{\sqrt{4-\pi}}$, and $\nu_{1,2} := \frac{\sqrt{4-\pi}(3-\pi \pm s)}{2(1 \pm s)}$. If $\phi \in(0,\pi)$ is such that
\begin{equation}\label{involvedcond}
 2\phi s\Theta_0^2 \left [2\phi s -\mu_1^2\mu_2^2 \left \{ \nu_1 \tanh\left (\textstyle{\frac12}\mu_1\phi\right )
 + \nu_2 \tanh\left (\textstyle{\frac12}\mu_2\phi\right ) \right \}\right ]^{-1} > 1\,,
\end{equation}
then $\:\sigma_\mathrm{disc}(H_{N,\phi}) \cap (-\infty,\Theta_0) \ne \emptyset$.
\end{theorem}

\smallskip

\noindent Numerical analysis of (\ref{involvedcond}) yields now the existence of at least one bound state for all $\phi \lesssim 0.583\pi$, which is an improvement with respect  to \cite{Bo03}. In principle one could proceed to larger values of $N$ but it becomes more difficult to execute the functional derivative optimization rigorously. Using the above Ansatz with $N=4$ we get \emph{numerically} $\phi \lesssim 0.595\pi$; we expect that the present method is not likely to work \emph{beyond $\phi \gtrsim 0.6\pi$}, still far from the `full' conjecture.

Finally, let us mention briefly a related problem, bound states of the Schr\"odinger operator with $\delta$ interaction on a broken line, now in presence of a homogeneous magnetic field. Here we can make the following claim \cite{ELP17}:

\begin{theorem} \label{thm:mgbroken}
Let $\phi \in (0,\pi)$ and $\beta > 0$ be fixed, and let further $F_{\phi,\beta}(\cdot)$ be defined by
$$
F_{\phi,\beta}(x,y) = 1 + \frac{x^4}{4} - x^2\Theta_{\delta,\beta} - \beta x \pi^{-1/2}e^{-y^2\tan^2(\phi/2)}
\left (1+\mathrm{erf}\,(y)\right )\,.
$$
If $\:\inf_{x,y\in (0,\infty)}F_{\phi,\beta}(x,y) < 0$, then $\:\sigma_\mathrm{disc}(H_{\delta,\phi},\beta) \cap (-\infty,\Theta_{\delta,\beta}) \ne \emptyset$.
\end{theorem}
\emph{Sketch of the proof:} We change the gauge, or equivalently, we rotate and shift the broken line $\Gamma$ by the angle $\pi/4-\phi/2$ counterclockwise and shift it by the vector $(-c,-c)^\top$, where $c > 0$ is a parameter to be determined. Using the trial function given in the polar coordinates $(r,\theta)$ by
$$
u(r,\theta) := e^{-ar^2/2}\,,\quad a > 0\,,
$$
we get the sought statement after analytical optimization with respect to the parameters $a, c > 0$. \hfill $\Box$

\smallskip

\begin{figure}
  \includegraphics[width=0.75\textwidth]{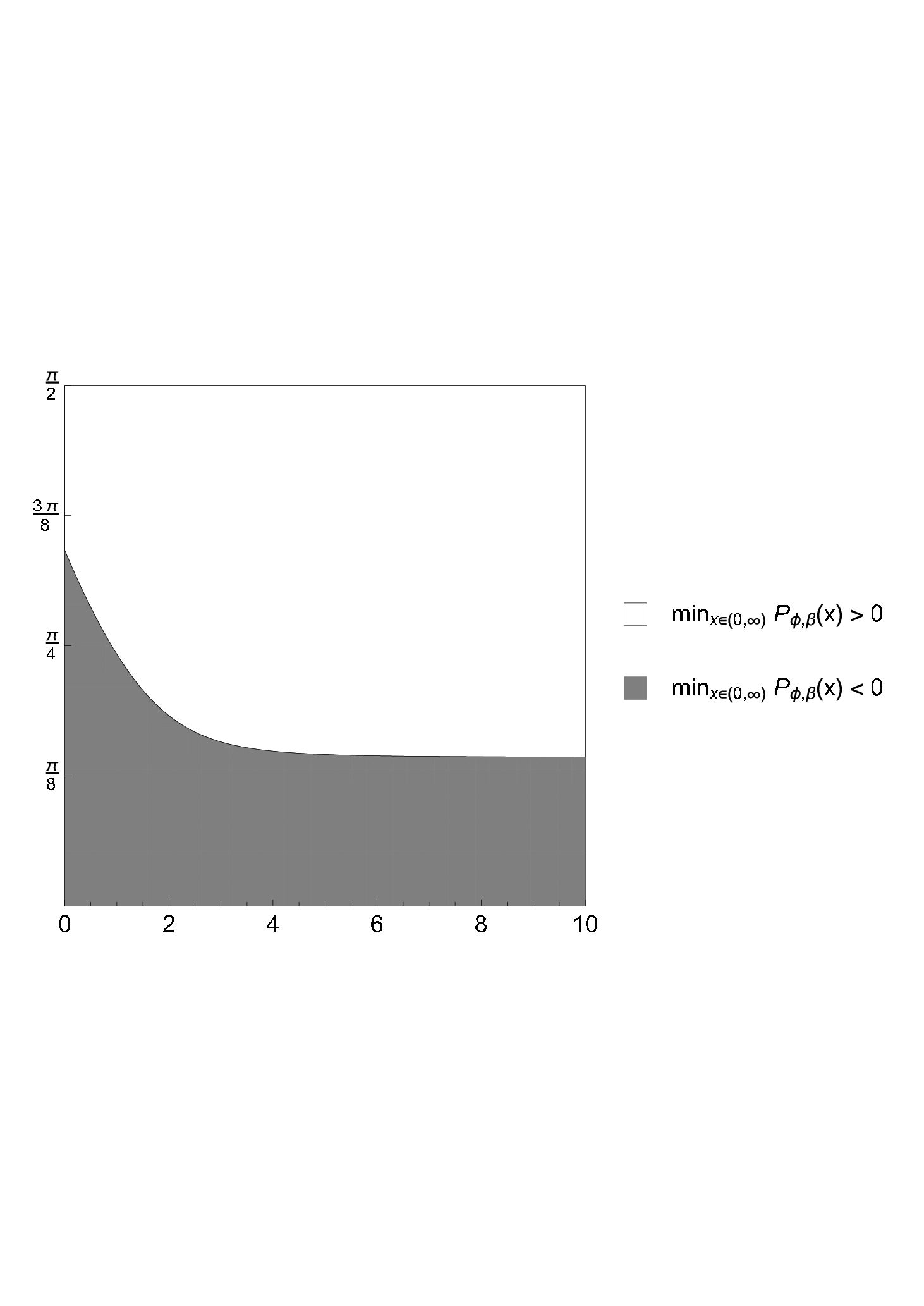}
\caption{The discrete spectrum of the Schr\"odinger operator refers to the shaded region}
\label{fig:broken}       
\end{figure}

A graphical solution to the above condition is shown on Fig.~\ref{fig:broken}. We see that bound states always exist provided the shape of $\Gamma$ is sharp enough, $\phi<\frac18\pi$. At the same time, the result expressed by Theorem~\ref{thm:mgbroken} is far from optimal. In particular, it is natural to expect that for weak magnetic fields (which by scaling correspond to large values of $\beta$) the bound states will survive even in situation close to the straight line, $\phi=\pi$.

\section{Conclusions}
\label{sec:concl}

Despite a number of results achieved in this area recently, many questions remain open. Let us mention briefly some of them:

\smallskip

\noindent -- The asymptotic expansions in Sec.~\ref{sec:strong} were derived under relatively strong regularity requirements to the interaction support $\Gamma$. It is obvious that they cannot hold, say, if the latter is not smooth; one is interested what could replace them is such situations. More generally, an open problem concerns the strong-coupling behavior in situations when $\Gamma$ has a more complicated topology, for instance, that of a \emph{branched graph.}

\smallskip

\noindent -- In particular, consider the operator $-\Delta_{\alpha,\Gamma}$ where $\Gamma$ is a graph with some edges semiinfinite and asymptotically straight and $\sigma_\mathrm{ess}(-\Delta_{\alpha,\Gamma}) = [-\frac14\alpha^2, \infty)$. As we have seen above in such situation we often have $\sigma_\mathrm{disc}(-\Delta_{\alpha,\Gamma})\ne \emptyset$ for $\alpha>0$. The natural question motivated by the analogous problem in quantum waveguides \cite{EK15} concerns the behavior of the `renormalized' operator $-\Delta_{\alpha,\Gamma} +\frac14\alpha^2$ as $\alpha\to\infty$, particular, the existence of the limit in the operator-norm sense. Motivated by \cite{ACF07,CE07,Gr08} we propose

\begin{conjecture}
The said limit, denoted as $-\Delta^\mathrm{ren}_{\Gamma}$, exists if $-\Delta_{\alpha,\Gamma}$ has a threshold resonance for some $\alpha>0$. It satisfies $\sigma_\mathrm{disc}(-\Delta^\mathrm{ren}_{\Gamma})=\emptyset$ but a nonempty discrete spectrum of the limiting operator could be achieved if we change simultaneously the geometry of $\Gamma$ in a suitable way.
\end{conjecture}

\smallskip

\noindent -- \emph{Scattering theory} for leaky wires have been so far worked out only in particular cases such as $\Gamma$ being a local deformation of a straight line in $\mathbb{R}^2\,$ \cite{EKo05}. For sufficiently smooth curves we expect that in the strong-coupling behavior of the scattering matrix the one-dimensional character will be dominant:

\begin{conjecture}
Let $\Gamma$ be a $C^4$-smooth and asymptotically straight planar curve. The on-shell scattering matrix $\Sigma_{\alpha, \Gamma}(k)$ at energy $k^2$ satisfies $\Sigma_{\alpha, \Gamma}(k-\frac14\alpha^2) \longrightarrow \Sigma_{S_\Gamma}(k)$ as $\alpha\to\infty$, where $\Sigma_{S_\Gamma}(k)$ is the on-shell scattering matrix referring to the one-dimensional comparison operator (\ref{compar2D}).
\end{conjecture}

\smallskip

\noindent -- Concerning Robin billiards, the only result known so far which makes it possible to distinguish individual eigenvalues is Theorem~\ref{thm:HK}. If the global maximum of the curvature is not unique, one has to consider tunneling between the corresponding `potential wells' in the way analogous to the multi-well problem in the usual Schr\"odinger operator theory \cite{HP15}. To extend the result of Theorem~\ref{thm:HK} to higher dimensions, one would have to address the known problem of frequency commensurability of the harmonic oscillator approximation at the bottom of the effective potential \cite{ADK16}.

\smallskip

\noindent -- \emph{Validity of the magnetic wedge conjecture} is another important open question, together with the analogous problem for \emph{magnetic Schr\"odinger operators} with a broken $\delta$ line (and more general supports). In the latter case we do not even have a numerical hint that motivates the conjecture in \cite{Ra17}; one could guess that a sufficiently strong magnetic field \emph{may destroy weakly bound states}.

\smallskip

\noindent -- Another open problem concerns for \emph{periodic manifolds}. One naturally expects that the spectrum of $-\Delta_{\alpha,\Gamma}$ with $\Gamma$ periodic in $d'$ directions, $0<d'<d$, would be absolutely continuous, but only a partial result is known, or a similar claim in the situation when the geometry of $\Gamma$ is trivial but the coupling strength is a periodic function, cf. \cite{EF07} and references therein.
\smallskip

\noindent -- The list may continue but we prefer to stop here with the hope that we managed to convince the reader that the subject of this survey is interesting and it offers still many challenges.

\begin{acknowledgements}
Our recent results discussed in this survey are the result of a common work with a number of colleagues, in the first place Jaroslav Dittrich, Sylwia Kondej, Christian K\"uhn, Vladimir Lotoreichik, Konstantin Pankrashkin, and Axel P\'erez-Obiol whom I am grateful for the pleasure of collaboration. The research was supported by the Czech Science Foundation (GA\v{C}R) within the project 17-01706S. Thanks also go to the referee for careful reading of the manuscript.
\end{acknowledgements}


\begin{thebibliography}{99}
%
%
\bibitem{ACF07}
S.A.~Albeverio, C.~Cacciapuoti, D.~Finco, Coupling in the singular limit of thin quantum waveguides, \emph{J. Math. Phys.} \textbf{48}, 032103 (2007)
\bibitem{AGHH}
S.~Albeverio, F.~Gesztesy, R.~H\o egh-Krohn, H.~Holden, \emph{Solvable Models in Quantum Mechanics}, 2nd edition; AMS Chelsea Publishing, Providence, R.I. (2005)
\bibitem{ADK16}
A.Yu.~Anikin, S.Yu.~Dobrokhotov, M.I.~Katsnel'son, Lower part of the spectrum for the two-dimensional Schr\"odinger
operators with periodic in one variable potential and applications to quantum dimers, \emph{Teoret. Mat. Fiz.} \textbf{188}, 288--317 (2016)
\bibitem{AGS87}
J.-P.~Antoine, F.~Gesztesy, J.~Shabani, Exactly solvable models of sphere interactions in quantum mechanics, \emph{J. Phys. A: Math. Gen.} \textbf{20}, 3687--3712 (1987)
\bibitem{BEL14}
J.~Behrndt, P.~Exner, V.~Lotoreichik, Schr\"odinger operators with $\delta$-interactions supported on conical surfaces, \emph{J. Phys. A: Math. Theor.} \textbf{47}, 355202 (2014)
\bibitem{BK13}
G. Berkolaiko, P. Kuchment, \emph{Introduction to Quantum Graphs}, $\,xiii+270\!$ p.; AMS, Providence, R.I. (2013).
\bibitem{Bo03}
V. Bonnaillie, Analyse math\'ematique de la supraconductivit\'{e} dans un domaine \'{a} coins: m\'ethodes semi-classiques et num\'eriques, \emph{Th\`ese de doctorat}, Universit\'{e} Paris XI, Orsay (2003)
\bibitem{CE07}
C.~Cacciapuoti, P.~Exner, Nontrivial edge coupling from a Dirichlet network squeezing: the case of a bent waveguide, \emph{J.~Phys.~A: Math.~Theor.} \textbf{40}, F511-F523 (2007)
\bibitem{DEKP16}
J.~Dittrich, P.~Exner, Ch.~K\"uhn, K.~Pankrashkin, On eigenvalue asymptotics for strong $\delta$-interactions supported by surfaces with boundaries, \emph{Asympt. Anal.} (2016) \textbf{97}, 1--25 (2016)
\bibitem{Ex08}
P.~Exner, Leaky quantum graphs: a review, in \emph{Proceedings of the Isaac Newton Institute programme ``Analysis on Graphs and Applications''}, AMS ``Proceedings of Symposia in Pure Mathematics'' Series, vol.~77, Providence, R.I.; pp.~523--564 (2008)
\bibitem{EF07}
P.~Exner, R.~Frank, Absolute continuity of the spectrum for periodically modulated leaky wires in $\mathbb{R}^3$, \emph{Ann. Henri Poincar\'{e}} \textbf{8}, 241--263 (2007)
\bibitem{EI01}
P.~Exner, T.~Ichinose, Geometrically induced spectrum in curved leaky wires, \emph{J. Phys. A: Math. Gen.} \textbf{34}, 1439--1450 (2001)
\bibitem{EJ13}
P.~Exner, M.~Jex, Spectral asymptotics of a strong $\delta'$ interaction on a planar loop, \emph{J. Phys. A: Math. Theor.} \textbf{46}, 345201 (2013)
\bibitem{EJ14}
P.~Exner, M.~Jex, Spectral asymptotics of a strong $\delta'$ interaction supported by a surface, \emph{Phys. Lett.} \textbf{A378}, 2091--2095 (2014)
\bibitem{EKo02}
P.~Exner, S.~Kondej, Curvature-induced bound states for a $\delta$ interaction supported by a curve in $\mathbb{R}^3$, \emph{Ann. H.~Poincar\'{e}} \textbf{3}, 967--981 (2002)
\bibitem{EKo03}
P.~Exner, S.~Kondej, Bound states due to a strong $\delta$ interaction supported by a curved surface, \emph{J. Phys. A: Math. Gen.} \textbf{36}, 443--457 (2003)
\bibitem{EKo05}
P.~Exner, S.~Kondej, Scattering by local deformations of a straight leaky wire, \emph{J. Phys. A: Math. Gen.} \textbf{38}, 4865-4874 (2005)
\bibitem{EKo08}
P.~Exner, S.~Kondej, Hiatus perturbation for a singular Schr\"odinger operator with an interaction supported by a curve in $\mathbb{R}^3$, \emph{J. Math. Phys.} \textbf{49}, 032111 (2008)
\bibitem{EKo15}
P.~Exner, S. Kondej: Gap asymptotics in a weakly bent leaky quantum wire, \emph{J. Phys. A: Math. Theor.} \textbf{48}, 495301 (2015)
\bibitem{EKL18}
P.~Exner, S.~Kondej, V.~Lotoreichik, Asymptotics of the bound state induced by $\delta$-interaction supported on a weakly deformed plane, \emph{J. Math. Phys.} \textbf{59}, 013051 (2018)
\bibitem{EK15}
P.~Exner and H.~Kova\v{r}\'{\i}k, \emph{Quantum Waveguides}, $\,xxii+382\!$ p.; Springer International, Heidelberg (2015)
\bibitem{ELP17}
P.~Exner, V.~Lotoreichik, A.~P\'erez-Obiol, On the bound states for magnetic Laplacians on wedges, \emph{Rep. Math. Phys.} (2018), to appear; \texttt{arXiv:1703.03667}
\bibitem{EM14}
P.~Exner, A.~Minakov, Curvature-induced bound states in Robin waveguides and their asymptotical properties, \emph{J. Math. Phys.} \textbf{55}, 122101 (2014)
\bibitem{EMP14}
P.~Exner, A.~Minakov, L.~Parnovski, Asymptotic eigenvalue estimates for a Robin problem with a large parameter, \emph{Portugal. Math.} \textbf{71}, 141--156 (2014)
\bibitem{EP14}
P.~Exner, K.~Pankrashkin, Strong coupling asymptotics for a singular Schr\"o\-din\-ger operator with an interaction supported by an open arc, \emph{Comm. PDE} \textbf{39}, 193--212 (2014)
\bibitem{ER16}
P.~Exner, J.~Rohleder, Generalized interactions supported on hypersurfaces, \emph{J. Math. Phys.} \textbf{57}, 041507 (2016)
\bibitem{EY02}
P.Exner, K.Yoshitomi, Asymptotics of eigenvalues of the Schr\"odinger operator with a strong $\delta$-interaction on a loop, \emph{J. Geom. Phys.} {\bf 41}, 344--358 (2002)
\bibitem{Fi17}
A.V.~Filinovskiy, On the asymptotic behavior of eigenvalues and eigenfunctions of the Robin problem with large parameter, \emph{J. Math. Model. Anal.} \textbf{22}, 37--51 (2017)
\bibitem{Gr08}
D.~Grieser,  Spectra of graph neighborhoods and scattering, \emph{Proc.~London Math.~Soc.} \textbf{97}, 718-752 (2008)
\bibitem{HK17}
B.~Helffer, A.~Kachmar: Eigenvalues for the Robin Laplacian in domains with variable curvature, \emph{Trans. AMS} \textbf{369}, 3253--3287 (2017)
\bibitem{HP15}
B.~Helffer, K.~Pankrashkin, Tunneling between corners for Robin Laplacians, \emph{J. London Math. Soc.} \textbf{91}, 225--248 (2015)
\bibitem{Ja01}
H.~Jadallah, The onset of superconductivity in a domain with a corner, \emph{J. Math. Phys.} \textbf{42}, 4101--4121 (2001)
\bibitem{JL16}
M.~Jex, V.~Lotoreichik, On absence of bound states for weakly attractive $\delta'$-interactions supported on non-closed curves in $\mathbb{R}^2$, \emph{J. Math. Phys.} \textbf{57}, 022101 (2016)
\bibitem{KL14}
S.~Kondej, V.~Lotoreichik, Weakly coupled bound state of 2-D Schr\"odinger operator with potential-measure, \emph{J. Math. Anal. Appl.} \textbf{420}, 1416--1438 (2014)
\bibitem{KP17}
H.~Kova\v{r}\'{\i}k, K.~Pankrashkin, On the p-Laplacian with Robin boundary conditions and boundary trace theorems, \emph{Calc. Var. PDE} \textbf{56}, 49 (2017)
\bibitem{LO16}
V.~Lotoreichik, T.~Ourmi\`eres-Bonafos, On the bound states of Schr\"odinger operators with $\delta$-interactions on conical surfaces, \emph{Comm. PDE} \textbf{41}, 999--1028 (2016)
\bibitem{OP18}
T.~Ourmi\`eres-Bonafos, K. Pankrashkin, Discrete spectrum of interactions concentrated near conical surfaces,
\emph{Appl. Anal.}, to appear; \texttt{arXiv:1612.01798}
\bibitem{PP16}
K.~Pankrashkin, N.~Popoff, An effective Hamiltonian for the eigenvalue asymptotics of the Robin Laplacian with a large parameter, \emph{J. Math. Pures Appl.} \textbf{106}, 615--650 (2016)
\bibitem{Po15}
N.~Popoff, The model magnetic Laplacian on wedges, \emph{J. Spect. Theory} \textbf{5}, 617--661 (2015)
\bibitem{Ra17}
N.~Raymond, \emph{Bound States of the Magnetic Schrödinger operator}, EMS Tracts in Mathematics, Z\"urich (2017)
\bibitem{RS}
M.~Reed, B.~Simon, \emph{Methods of Modern Mathematical Physics, IV.~Analysis of Operators}, Academic Press, New York (1978)


\end{thebibliography}


\end{document}